\documentclass[aps,twocolumn]{revtex4-2}  
\usepackage [cp1251]{inputenc}
\usepackage{amsmath}
\usepackage{bm}
\usepackage{graphicx}
\usepackage{bm}
\usepackage{epsfig}
\usepackage{epstopdf}
\usepackage{amsmath}
\usepackage{color}

\begin{document}
\title{The lock-on effect and collapsing bipolar Gunn domains in high-voltage GaAs avalanche \textit{p-n} junction diode}

\author{A. Rozkhov, M. Ivanov, P. Rodin$^{*}$}

\affiliation{Ioffe Institute, 194021 St.~Petersburg, Russia\\
$^{*}$rodin@mail.ioffe.ru }

\begin{abstract}
We present experimental evidence and physics-based simulations of the lock-on effect in high-voltage GaAs avalanche diodes. The avalanche triggering is initiated by steep voltage ramp applied to the diode and in-series 50 $\Omega$ load. After subnanosecond avalanche switching the reversely biased GaAs diode remains in the conducting state for the whole duration of the applied pulse (dozens of nanoseconds). There is no indication of the $p-n$ junction recovery that is commonly expected to develop on the nanosecond scale due to the drift extraction of non-equilibrium carriers. The diode voltage keeps a constant value of $\sim$70 V much lower than the stationary breakdown voltage of 400 V. Numerical simulations reveal that the conducting state is supported by impact ionization in narrow high-field collapsing Gunn domains as well as in quasi-stationary cathode and anode ionizing domains. Collapsing Gunn domains spontaneously appear in the dense electron-hole plasma due to the negative differential mobility of electrons in GaAs. The effect resembles the lock-on effect of GaAs bulk photoconductive switches but is observed in reversely biased $p-n$ junction diode switched by a non-optical method.
\end{abstract}

\date{June 13, 2022}
  \maketitle

\section{Introduction}
The lock-on effect, originally found in bulk GaAs optically activated switches \cite{Loubriel,Zutavern87,Zutavern90}, manifests itself in the conductive state that persists for virtually unlimited time period after the termination of the triggering optical pulse \cite{Loubriel,Zutavern87,Zutavern90,Rosen,ZutavernBook}. The physical mechanism of the lock-on in GaAs, which had been a mystery for decades, was eventually attributed to narrow ionizing Gunn domains --- coined collapsing field domains (CFDs) \cite{Hu,Chowdhury}. Discovered in subnanosecond range GaAs avalanche transistors \cite{Vainshtein05}, CFDs together with avalanche injection have been employed to explain the lock-on effect in GaAs optically activated switches \cite{Hu,Chowdhury}. CFDs spontaneously appear in the dense electron-hole plasma due to the negative differential mobility of electrons in GaAs \cite{Hu,Chowdhury,Vainshtein05} but are strikingly different from both classical Gunn domains in n-type GaAs \cite{Gunn,Kroemer} and non-ionizing bipolar Gunn domains studied in 70-ties \cite{Gelmont1,Gelmont2}. 

In this paper we report the experimental observation of the lock-on effect in GaAs \textit{p-n} junction diodes triggered in the avalanche mode by a non-optical method. Avalanche switching of 0.4 kV GaAs $p^+-p^0-i-n^0-n^+$ structures is initiated by the application of triggering pulse with steep ($\sim$1.5 kV/ns) voltage ramp and reverse polarity. The triggering pulse length exceeds 40 ns. Despite the reverse diode bias, the conducting state endures for the whole duration of the applied pulse. The residual voltage of 70 V is much smaller than the stationary breakdown voltage. Light emission indicates the current flow in narrow surface conducting channels. Simulations of the inner device dynamics reveal the appearance of ionizing traveling CFDs as well as formation of quasi-stationary ''cathode'' and ''anode'' high-field ionizing domains located at the $p^+-p^0$ and $n^+-n^0$ junctions, respectively. We demonstrate that impact ionization within these narrow ($\sim$2 {\textmu}m) high-field domains keeps electron-hole plasma concentration of $10^{17} \text{cm}^{-3}$ in the conducting channel and hence supports the conducting state of the reversely biased GaAs diode.

 \section{Structures and experimental setup} \label{setup}
 \begin{figure*}[ht] 	
 	\centering
 		\includegraphics[width=14 cm]{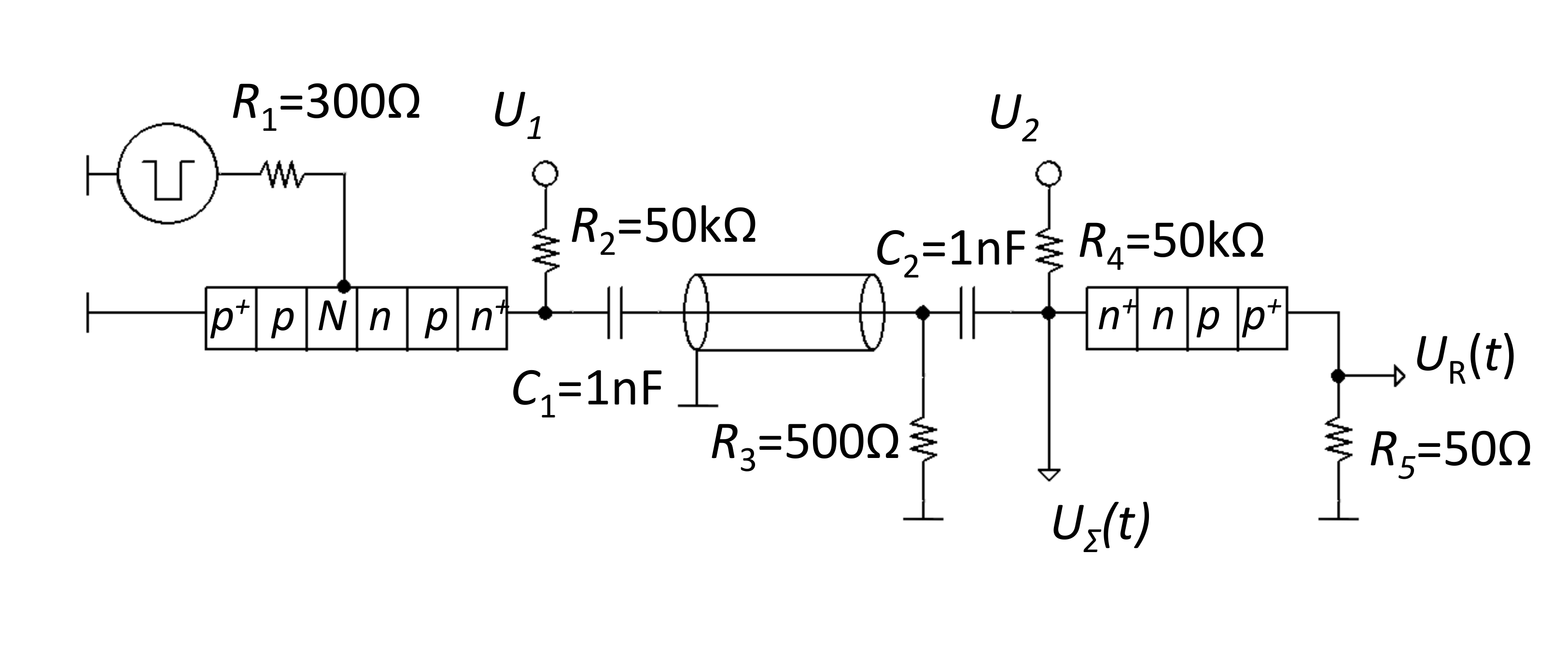}
 	\caption{Sketch of the experimental setup.} \label{geometry}
 \end{figure*}

GaAs $p^+-p^0-i-n^0-n^+$ diodes with the stationary breakdown voltage of $U_b \approx 400$ V and diameter 0.5 mm have been fabricated by liquid phase epitaxy on Zn-doped substrate. The width of the low-doped $p^0-i-n^0$ region is $\sim$45 {\textmu}m. The recombination lifetime of nonequilibrium carriers measured by the Lax method is about 100 ns. 

The test bench is sketched in Fig.~1. The triggering generator is based on the GaAs-AlGaAs heterojunction $p^+-p-N-n-p-n^+$ photo-injection thyristor \cite{Korolkov} that is capable to form pulses with rise-time $\sim$300 ps on the matched 50 $\Omega$ load. The pulse amplitude up to 400 V is tunable by the voltage $U_1$. As the capacitor $C_1$ discharges this amplitude decays exponentially with characteristic time $\tau_p=RC_1\approx 50$ ns  that exceeds the drift extraction time ($\sim$1 ns) and is comparable to the non-radiative recombination time ($\sim$100 ns) [see the inset in Fig.~2a]. The triggering pulses are applied to the GaAs $p^+-p^0-i-n^0-n^+$ diode under investigation and in-series 50 $\Omega$ load via the coaxial transmission line with repetition frequency 1 KHz. Prior to the application of the triggering pulse the diode is reversely biased to the voltage $U_2$ (Fig.~1). Two-channel oscilloscope connected via wideband attenuator lines measures the voltage $U_R(t)$ across the 50 $\Omega$ load (channel 1). The voltage $U_\Sigma (t)$ across the diode and the load is measured by means of a high voltage probe (channel 2). These measurements determine the current $I(t)=U_R(t)/R$ and the device voltage $U_D(t)=U_\Sigma (t)-U_R(t)$. The time resolution is about 300 ps.
\newline

\section{Experimental results} \label{ExpRes}
The measured device voltage $U_D(t)$ and current $I(t)$ are shown in Fig.~2 (green, blue and red lines for $U_2=$100, 200, 300 V, respectively) together with the triggering generator pulse measured separately on the matched 50 $\Omega$ load (the inset in Fig.~2a). In the initial stage the voltage at the anode nearly doubles with respect to the triggering pulse amplitude due to the reflection of the incident wave from the non-conducting diode in the transmission line \cite{Ivanov21,Ivanov22}. The voltage $U_D$ across the diode increases and eventually exceeds the stationary breakdown voltage $U_b$. This leads to the subnanosecond avalanche transient to the conducting state (t $\approx$ 5.5 ns). The switching transient develops in the so-called delayed impact ionization mode that is well-known for both Si and GaAs diodes (see \cite{Ivanov22,Brylevskiy17,Brylevskiy16} and references therein) and used in pulse power applications \cite{Kardo,Focia,Gusev,Jiang,Rukin,Merensky,Kesar,Huang}. After avalanche switching the current reaches the value of 6-10 A that increases with the initial bias $U_2$. The device voltage $U_D$ drops to the residual value of  $\sim$370 V much smaller than $U_b$. In this paper, we skip the details of the avalanche switching transient and focus on the post-breakdown dynamics. 

The common concept of the post-breakdown dynamics assumes that the drift extraction of non-equilibrium carriers \cite{Benda} eventually leads to the recovery of the reversely biased p-n junction. Alternatively, for high amplitude and long duration of the applied pulse the diode may undergo a transition to the high-voltage double injection mode \cite{Levinshtein,Ivanov20}. Both scenarios imply that the device voltage increases from low residual value to high values close to the stationary breakdown voltage $U_b$ within nanoseconds. However, our experimental results are in strong contradiction to this widely adopted view.

Indeed, after avalanche switching the diode remains in the conducting state for dozens of nanoseconds (about 40 ns in our experiments) without any indication of the $p-n$ junction recovery (Fig.~2). The diode voltage does not change with time and keeps the residual value of $\sim$70 V acquired after avalanche switching, much lower than stationary breakdown voltage $U_b=400$ V (Fig.~2a). The current decreases with time only because the applied voltage decreases as the capacitor $C_1$ discharges (Fig.~2b). The current is determined mostly by the 50 $\Omega$ external load and the applied voltage. The time period of $\sim$40 ns is much longer than drift extraction time and comparable to the non-radiative recombination time of 100 ns. Hence we conclude that the reversely biased diode GaAs exhibits lock-on in the conducting state. Preliminary observation of light emission performed by near-infrared low-frequency camera indicates the formation of narrow surface conducting channels. 
\newline

\section{Simulations} \label{Simulations}
 Device simulations reveal the physical mechanism of the experimentally observed lock-on phenomena. The drift-diffusion transport equations have been solved together with the Poisson equation using TCAD Silvaco. We use approximation
\begin{equation}
	{v}_n(E)=\frac{{\mu }_nE+{\nu }_s{\left(E/E_t\right)}^4}{1+{\left(E/E_t\right)}^4}\times \left[a+b\times \mathrm{exp}(-E/E_0)\right]
\end{equation}  
from Ref. 8 for the electron drift velocity $v_n(E)$ with parameters $E_t$ = 4 kV/cm, $v_s$ = 9.5$\times10^6$ cm/s, a = 0.576, b = 0.49, $E_0=1.5\times 10^5$ V/cm. This approximation describes negative differential mobility in the whole range of relevant electric fields and represents a simplified version of the more elaborated dependence suggested later in Ref. \cite{Vainshtein08}. We adopt standard approximations for impact ionization coefficients 
\begin{equation}
	{\alpha }_{n,p}={\alpha }^{\infty }_{n,p}\times \mathrm{exp}\left[{-\left({E^{\mathrm{crit}}_{n,p}}/{E}\right)}^{{\beta }_{n,p}}\right]
\end{equation}
with the following parameters: $\beta_n=1.82$, $\beta_p=1.75$, $\alpha_n^\infty=1.889\times10^5$ $\text{cm}^{-1}$, $\alpha_p^\infty=2.215\times10^5$ $\text{cm}^{-1}$, $E_n^{crit}=5.75\times10^5$ V/cm, $E_p^{crit}=6.57\times10^5$ V/cm \cite{Vainshtein08}. Non-radiative, radiative and Auger recombination processes are taken into account according to the standard Silvaco routine for GaAs \cite{Vainshtein08}. Spatial inhomogeneity of the avalanche switching is modelled by dividing the device cross-section S into two 
parts: the active part $S_a= S/K$ and passive part $S_p=S-S_a$ where the impact ionization is artificially switched off, as it has been done in Ref. \cite{Ivanov22}. These active and passive parts are treated as two parallel diodes. The parameter $K=S/S_a$ serves as a fit parameter. The module ''Transmission Line'' of TCAD Silvaco has been used to account for the effects related to the propagation of the triggering generator pulse in the transmission line and its reflection from the avalanche diode.

 \begin{figure}[] 	
 	\centering
 		\includegraphics[width=7.5 cm]{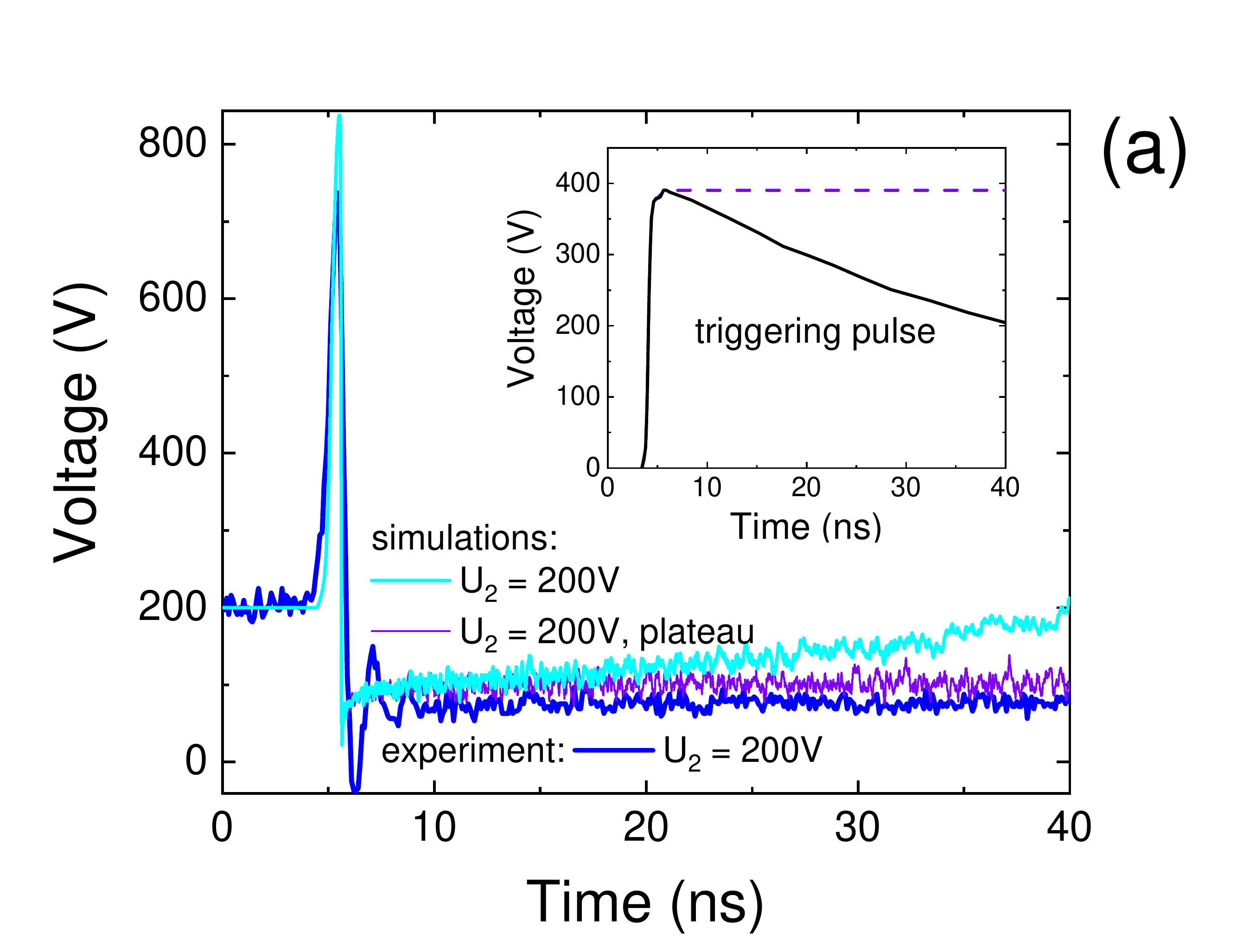}
		\includegraphics[width=7.5 cm]{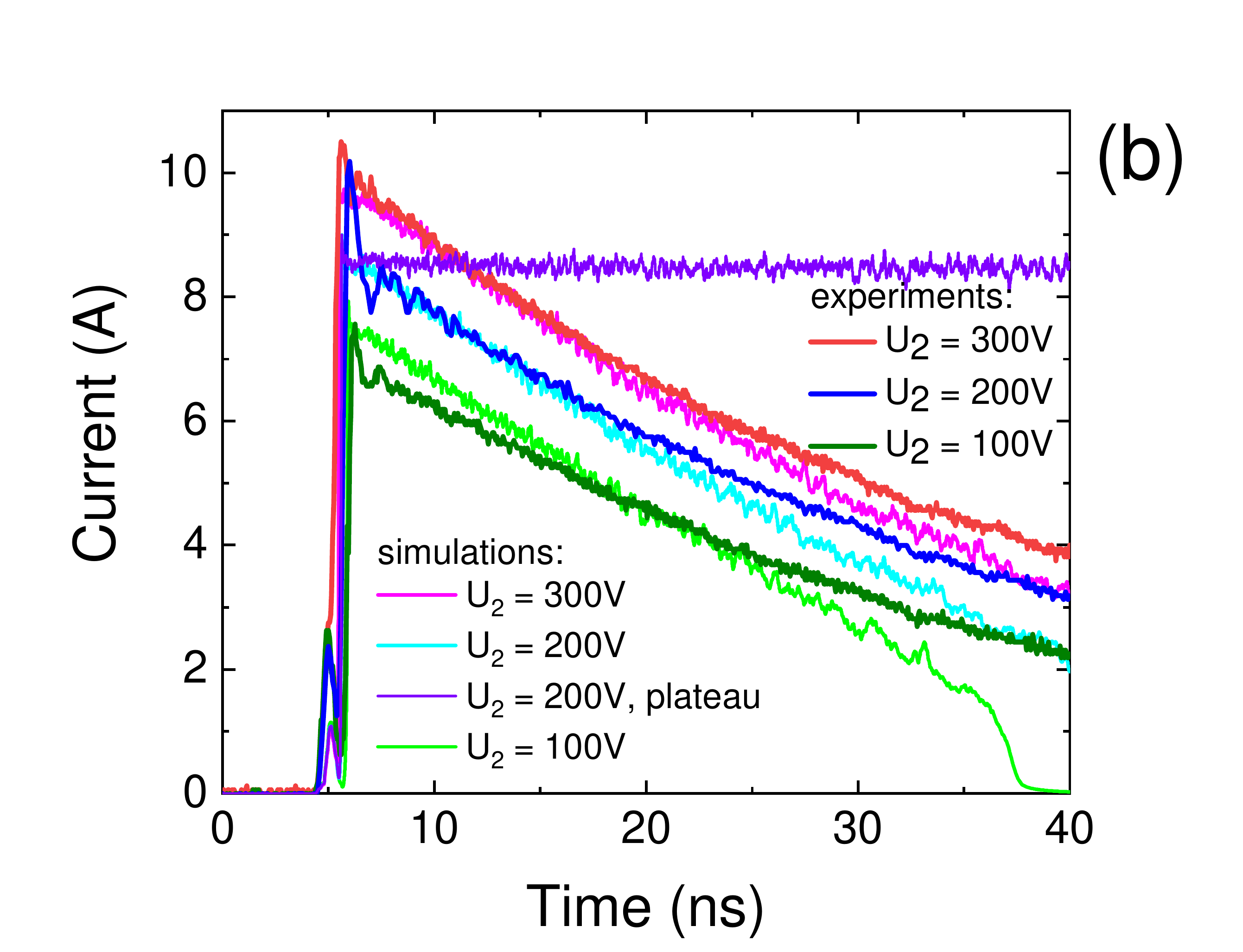}
 	\caption{Device voltage and current during subnanosecond avalanche switching (from 5 to 6 ns) and on the following lock-on stage (from 5 to 40 ns). The results of measurement and simulations are shown for the initial reverse bias $U_2$ = 200 V in panel (a) and three different values $U_2$ = 100, 200 and 300 V in panel (b). The inset in panel (a) shows the triggering pulse (solid line). The violet curves show the voltage and current for the quasi-rectangular triggering pulse shown by the dashed violet line in the inset [panel(a)] and initial bias $U_2$ = 200 V. } \label{2}
 \end{figure} 

The simulated dependencies $I(t)$ and $U_D(t)$ are shown in Fig.~2 (light green, cyan and magenta lines for $U_2=$100, 200 and 300 V, respectively). The agreement with experiments is achieved under the assumption that a small fraction of the device cross-section is switched-on. The presented results correspond to K=100. Reasonable agreement is achieved in the range of K from 50 to 150. Note that avalanche switching in narrow conducting channels is typical for superfast avalanche diodes [27,31]. 
The results of simulations reproduce both avalanche switching and the lock-on state. The calculated device voltage in the lock-on state is about 120 V vs 70 V in the experiment (Fig.~2a). The calculated device voltage slowly increases with time whereas the measured one keeps a constant value (Fig.~2a). There is also qualitative discrepancy between experiments and simulations at large time for the initial biases $U_1=100$ V: the simulations predict the $p-n$ junction recovery which is not observed in the experiment. We attribute both discrepancies to the assumption that the cross-section of the conduction channel is time-independent and will discuss these discrepancies in details in Sec. 6. The current density in the conducting channel is evaluated as $\sim$500 kA/$\text{cm}^2$ for $K=100$. Adiabatic Joule self heating of the conducting channel results in the temperature rise of 5 K/ns and hence may restrict the repetition rate. However, the temperature rise during a single pulse is not critical.

 \begin{figure}[] 	
 	\centering
 		\includegraphics[width=7.5 cm]{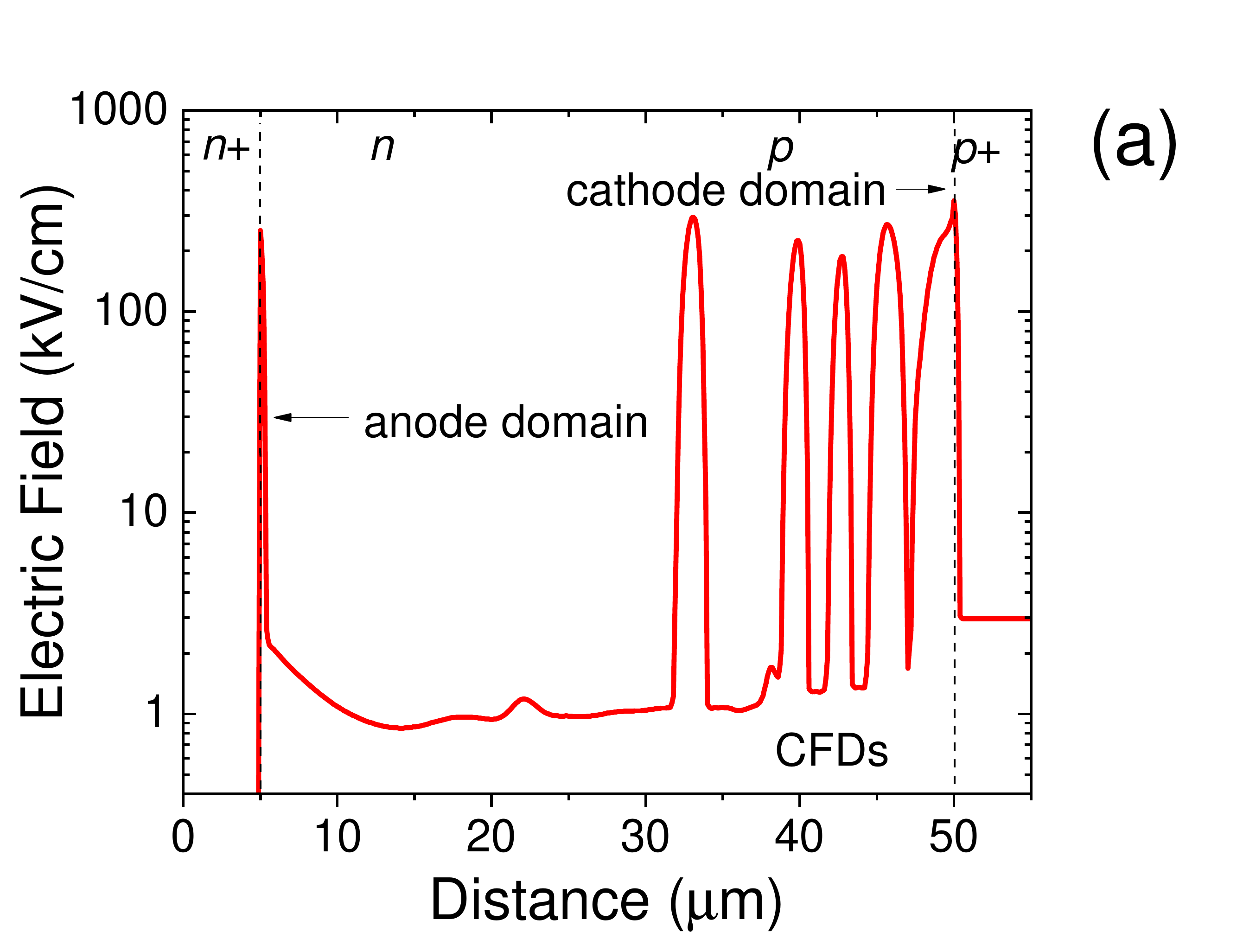}
		\includegraphics[width=7.5 cm]{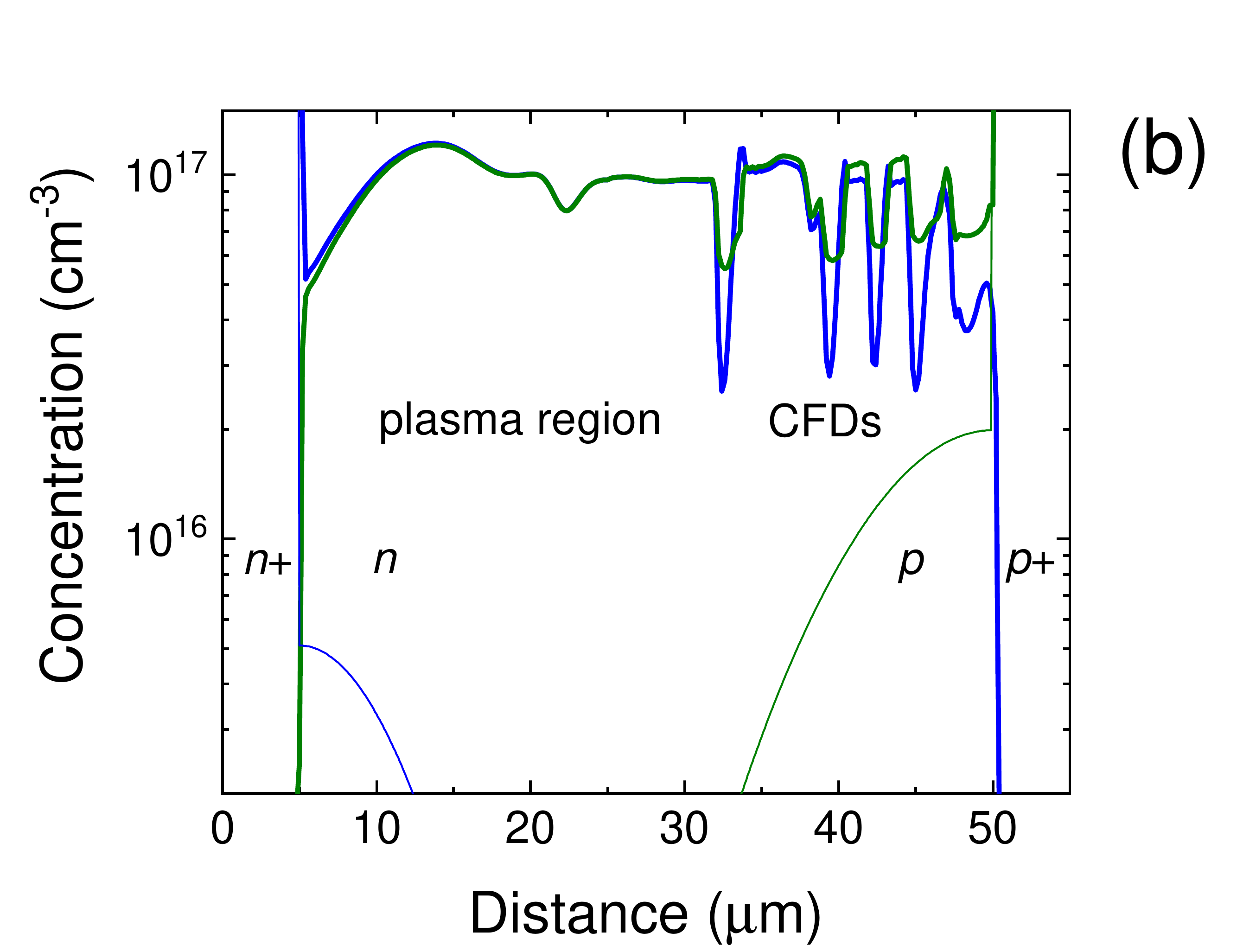}
 	\caption{Distribution of electric field $E(x,t)$ [panel (a)], electron and hole concentrations $n(x,t)$, $p(x,t)$ [blue and green lines in panel (b), respectively] in the conducting channel on the lock-on stage at $t=31.072$ ns. Thin blue and green lines in panel (b) depict the net doping of the $p^+-p^0-i-n^0-n^+$ diode.} \label{3}
 \end{figure} 

In our experiment, the phenomenon of the self-supporting conducting state is somewhat blurred by the decrease of the applied voltage that leads to a decrease of total current and hence current density. To justify the existence of the quasi-stationary conducting state we have performed simulations with quasi-rectangular pulse (see the inset in Fig.~2a, dashed violet line) and the initial bias $U_2=200$ V. The results are shown in Fig.~2 (violet curves). As seen, for constant applied voltage the device current and voltage are time-independent. The lifetime of the conducting state is limited only by Joule self-heating and related thermal effects.

\section{Collapsing field domains and the lock-on mechanism} \label{Domains}

\begin{figure}[] 	
 	\centering
 		\includegraphics[width=7.5 cm]{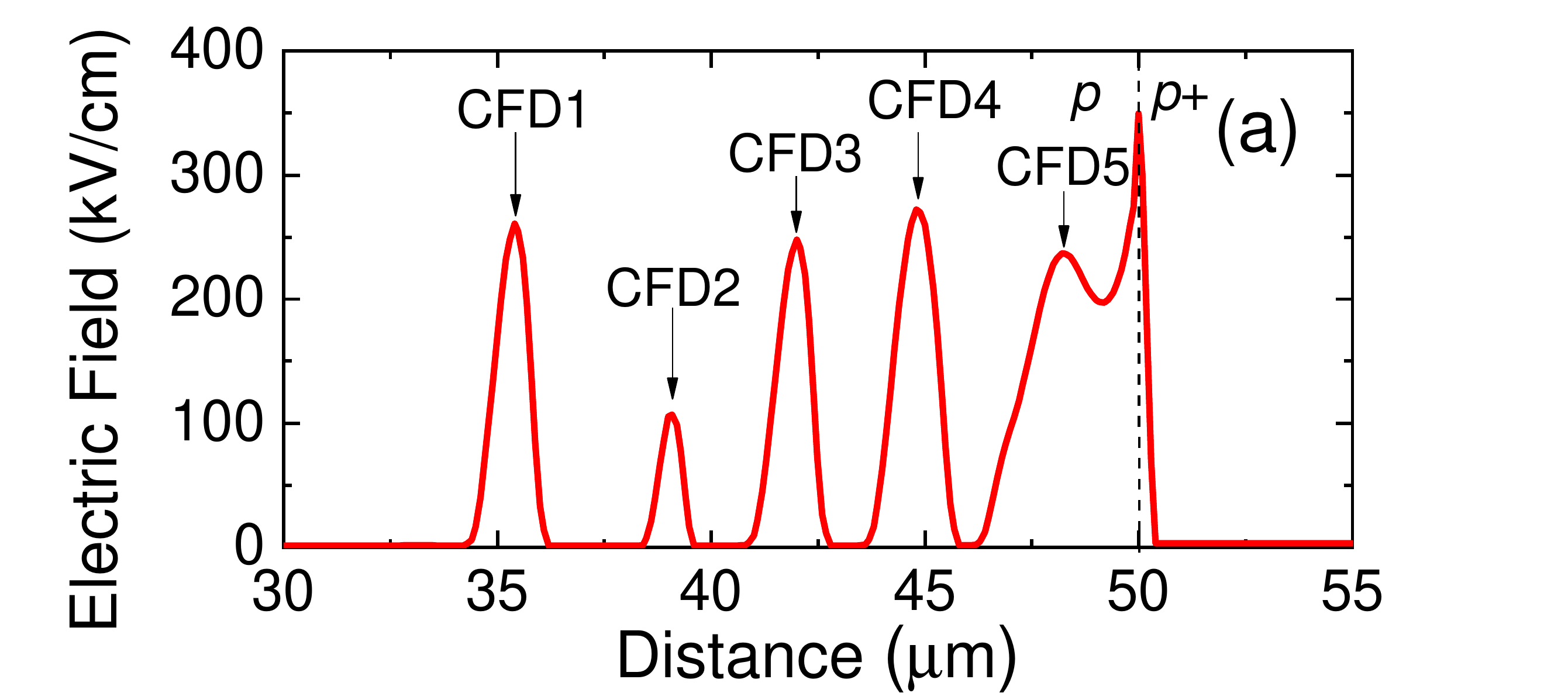}
		\includegraphics[width=7.5 cm]{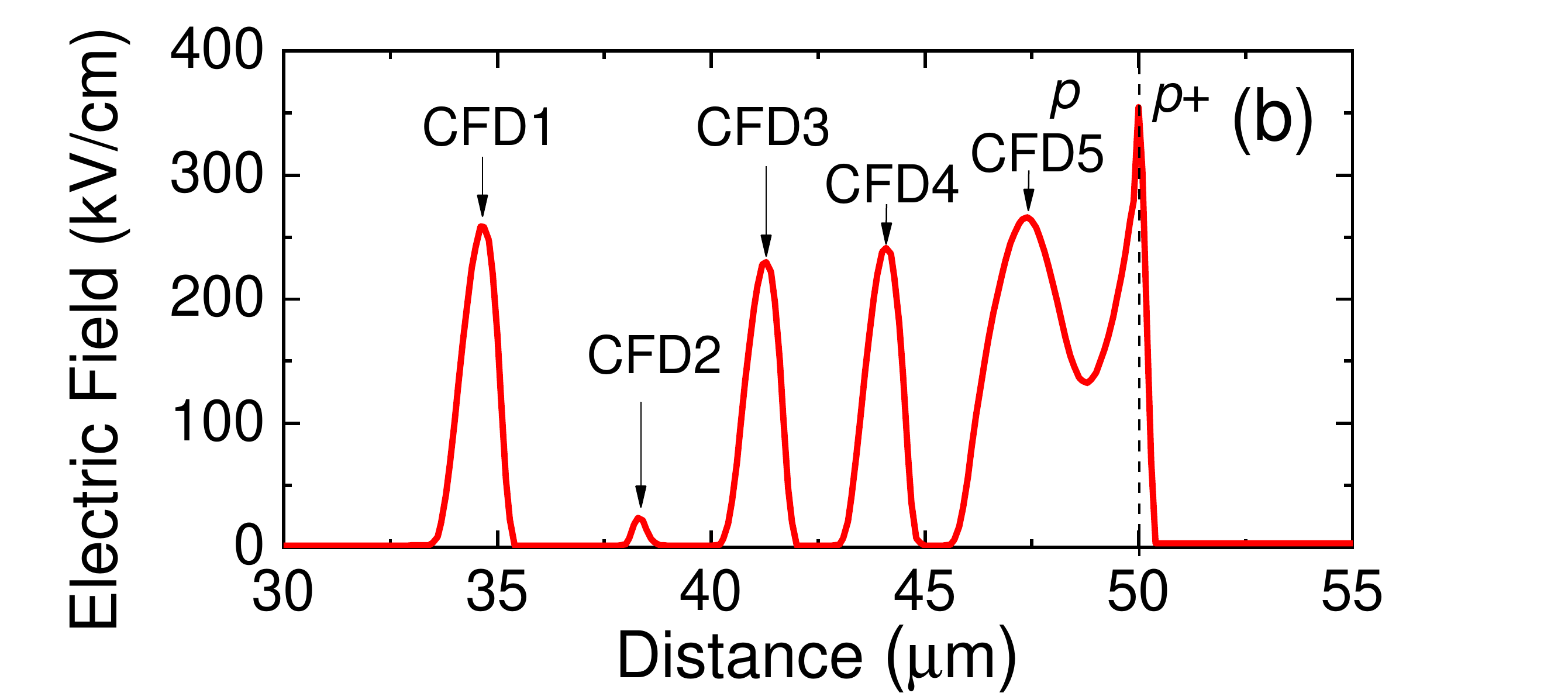}
		\includegraphics[width=7.5 cm]{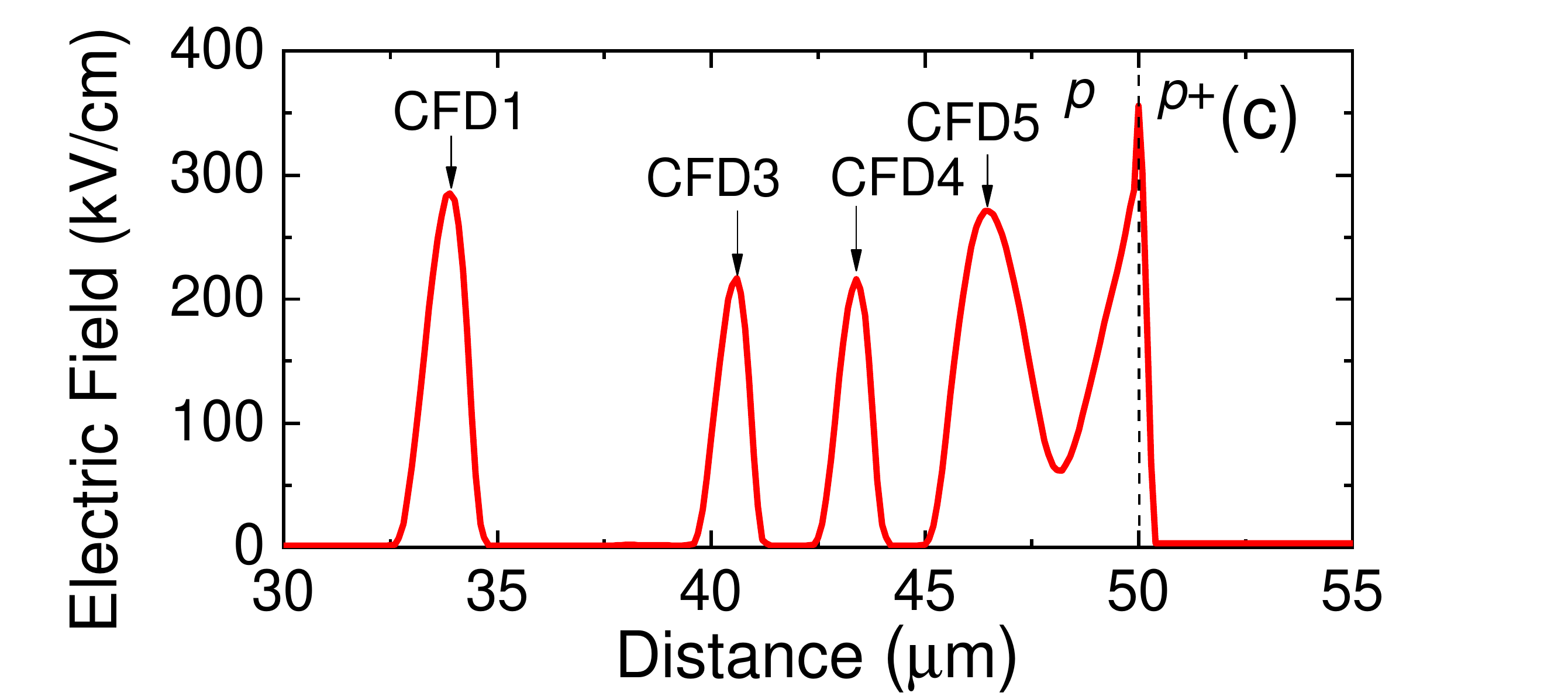}
		\includegraphics[width=7.5 cm]{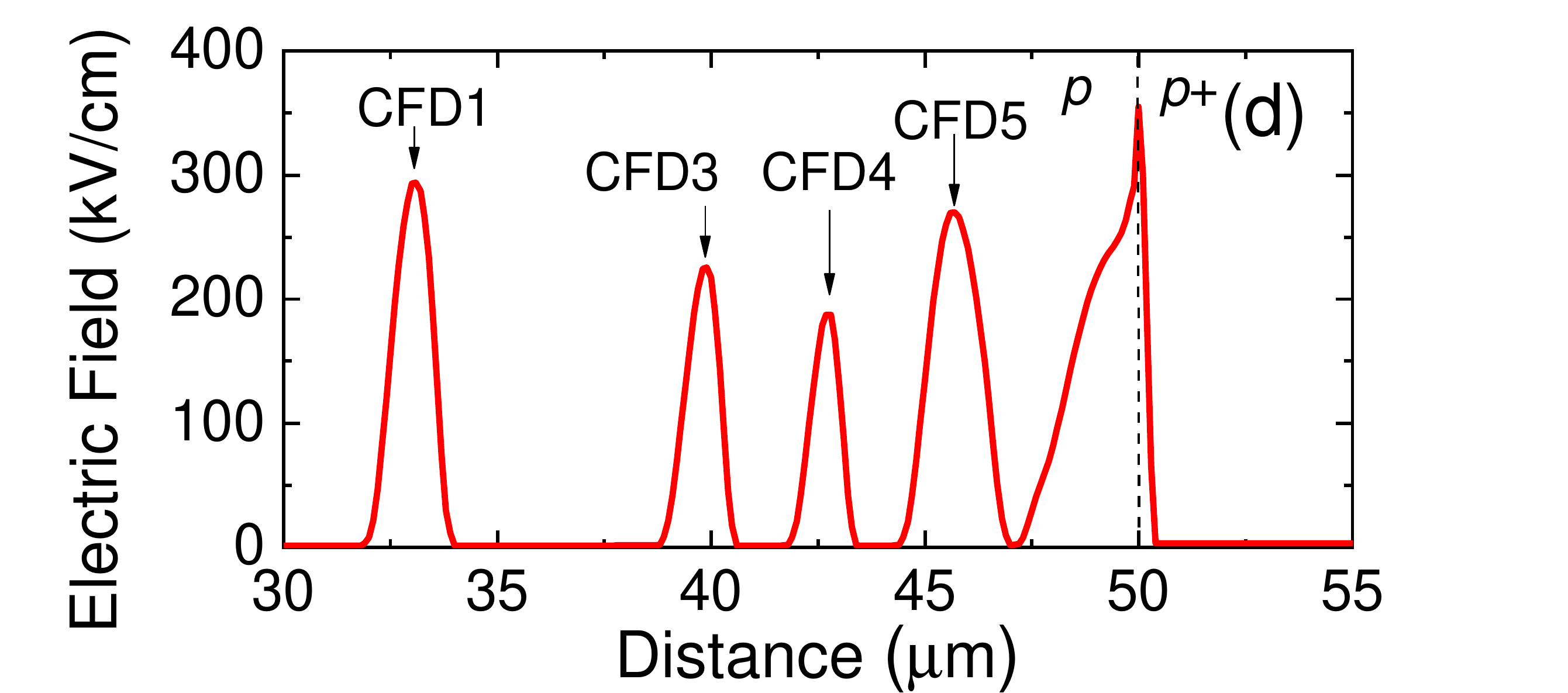}
 	\caption{Spatio-temporal dynamics of traveling collapsing field domains (CFDs) and the cathode domain during the lock-on. Distributions of electric field $E(x,t)$ are shown with 8 ps time step at the subsequent moments time moments $t=$ 31.048 (a), 31.056 (b), 31.064 (c), 31.072 ns (d).} \label{3}
 \end{figure} 

The electric field and the concentration profiles at the time moment $t=31.072$ ns ($\sim$26 ns after the avalanche switching) is shown in Fig.~3. The average concentration of non-equilibrium electron-hole plasma in the conducting channel is about $10^{17} \text{cm}^{-3}$. This value exceeds the dopant concentration by orders of magnitude. High irregular spikes of electric field correspond to the traveling bipolar Gunn domains --- CFDs (Fig.~3a). The electric field in CFDs is sufficient for intense impact ionization. Apart from CFDs, there are two steady ionizing domains at the edges of $p^+-p^0$ and $n^+-n^0$ regions --- cathode and anode domains, respectively. The time-dependent dynamics of the CFDs is illustrated in Fig.~4. The anode domain is a stationary one, the cathode domain ''breathes'' with time as traveling CFDs split from it one by one (see Fig.~4). The CFDs travel across the $p^0$-layer and most of them collapse before reaching the $n^0$-layer. The loss of nonequilibrium carriers due to the drift extraction and recombination is compensated by impact ionization in these narrow high-field domains despite the fact that the average electric field $\sim$10 kV/cm is much lower than the impact ionization threshold (~$\sim$250 kV/cm). 
Each CFD is characterized by the narrow width (about 2 $\mu$m) and the strong ionizing electric field (about 300 kV/cm). The CFDs drift opposite to the electric field with velocity of $\sim10^7$ cm/s (Fig.~4). These features correspond to the results of previous simulations of CFDs in avalanche bipolar transistors \cite{Vainshtein05,Vainshtein08} and photoconductive switches \cite{Hu,Chowdhury}. In contrast to the classical monopolar Gunn domains in n-type GaAs \cite{Gunn,Kroemer}, CFDs do not keep their shape and velocity: they appear and then collapse in an irregular way (Fig.~4). Traveling domains are generated at the edge of the cathode domain near the $p^+-p^0$ interface and cease in the middle of the low-doped $p^0-i-n^0$ region. Such irregular spatio-temporal dynamics can be characterized as deterministic spatio-temporal chaos \cite{Schoell}. Irregular domain generation induces stochastic current oscillation in the external circuit with GHz-range frequencies according to the simulation results. Note that sub-terahertz oscillations from bipolar avalanche transistors exhibiting bipolar CFDs, originally reported in Ref. \cite{Vainshtein07}, are nowadays used in imaging systems \cite{Vainshtein18,Vainshtein19}. The question of whether the observed lock-on effect in the GaAs diode is associated with sub-terahertz emission is currently under investigation.

\section{Discussion} \label{Discussion}
The lock-on in GaAs avalanche $p-n$ junction diodes reported here bears strong similarities to the lock-on effect in optically activated nonlinear switches based on semi-insulating GaAs \cite{Loubriel,Zutavern87,Zutavern90,Rosen,ZutavernBook,Hu,Chowdhury}. Although the initial switching mechanism, geometry and doping are drastically different, in both cases (i) the device maintains its conductivity for a virtually unlimited period of time after switching transient from blocking to a conductive state, (ii) the average residual electric field exceeds the 4 kV/cm and corresponds to negative differential mobility of electrons in GaAs, (iii) the current flows in narrow channels (or filaments) which ensures high current density. The simulations presented here and previous simulations of photoconductive switches \cite{Hu,Chowdhury} indicate the universal mechanism of lock-on effect: impact ionization in the narrow anode and cathode domains and in collapsing field domains that appear and propagate in the dense electron-hole plasma due to the bipolar Gunn effect. 

Let us address the main discrepancies between the experiment and simulations. First, the measured device voltage keeps constant value whereas the simulated one rises (Fig.~2a). Second, the simulated current is significantly smaller than the measured one at end of the pulse (Fig.~2b). For the lowest value of the initial bias $U_2=100$ V this difference results in qualitative discrepancy: the $p-n$ junction recovery is predicted by simulations at $t\approx$ 38 ns (Fig.~2b, the light green line). These two discrepancies have a common origin: the simulations underestimate the current density at large times. Indeed, in simulations we assume that the total cross-section of the conducting channels does not change in time: $S_a=S/K=$const since $K=$const. However, there are several mechanisms capable to change the cross-section and even the number of conducting channels during a single pulse. These mechanisms are recombination and internal photoionization, the latter often coined photon recycling \cite{Chowdhury}. Predomination of recombination results in shrinking of the channel and increase of the current density. Predomination of photon recycling results in channel broadening and decrease of the current density \cite{Vainshtein07}. Both shrinking and broadening are beyond our model. We propose that in experiments the total cross-section of the conducting channels decreases with time. This partly compensates for the decrease of the current density caused by the falling amplitude of the triggering pulse. Remarkably, for the applied pulse with plateau (inset in Fig.~2a, dashed violet line) the simulations predict a steady lock-on state with time-independent device voltage and current (Fig.~2, violet curves).
Ionization of deep impurities is known to play a significant role in GaAs switching devices \cite{Prudaev}. This process is also not accounted for in our simulations. The discrepancy of measured and simulated current on the early stage of the switching transient (interval from 4.5 to 5.5 ns in Fig.~2b, so-called displacement current) may be caused by redistribution of the electric charge in the reversely biased $p-n$ junction due to ionization of deep impurities. 

\section{Conclusion} \label{Conclusion}
In conclusion, high-voltage GaAs $p-n$ junction diodes switched in the avalanche mode by application of steep reverse voltage ramp exhibit lock-on in the conducting state for the period that is determined by the external circuit and exceeds 40 ns. This time period is much longer than drift extraction and comparable to the recombination time. During the lock-on, the reverse voltage across the diode with stationary breakdown voltage $\sim$0.4 kV is about 70 V which corresponds to the average electric field of 10 kV/cm. The current transport occurs in narrow surface luminous channels. Numerical simulations indicate that a possible mechanism that supports conductivity in the low average electric field is related to the bipolar Gunn effect. Electron-hole plasma is generated within collapsing field domains --- micrometer-size traveling regions of strong electric field --- and steady narrow cathode and anode domains at the $p^+-p^0$ and $n^+-n^0$ junction. Impact ionization processes in these narrow domains support electron-hole plasma concentration. Chaotic spatio-temporal dynamics of collapsing field domains generates high-frequency oscillations in the external circuit. 

\section*{Acknowledgments}
 We are grateful to V.I. Brylevskiy for helpful discussions.

\end{document}